\documentclass[lettersize,journal]{IEEEtran} % Formatting as required by IEEE
\usepackage{graphicx} % Required for inserting images
\usepackage{lipsum}
\usepackage{braket} % To write quantum states
\usepackage{float} % Omits 'float' error when captioning
\usepackage{caption} % Allows captioning
\usepackage{subcaption} % Allows subcaptioning
\usepackage{comment} % M ultiline commenting
\usepackage{tabularx} % Provides the 'X' type column for better looking tables
\usepackage{times}

\usepackage{cite}
\usepackage{amsmath,amssymb,amsfonts}
\usepackage{algorithmic}
\usepackage{xcolor}
\usepackage{textcomp}
\IEEEoverridecommandlockouts
\IEEEaftertitletext{\vspace{-1\baselineskip}}
\def\BibTeX{{\rm B\kern-.05em{\sc i\kern-.025em b}\kern-.08em
    T\kern-.1667em\lower.7ex\hbox{E}\kern-.125emX}}

\title{Training microwave pulses using quantum machine learning}

\author{Jaden Nola, Uriah Sanchez, Anusha Krishna Murthy, Elizabeth Behrman, and James Steck
\thanks{J. Nola and E. Behrman are with the Department of Mathematics, Statistics, and Physics, Wichita State University, Wichita, KS 67260-0033 USA; email jxnola1@shockers.wichita.edu; elizabeth.behrman@wichita.edu} 
\thanks{U. Sanchez is with the Department of Biomedical Engineering, Wichita State University, Wichita, KS 67260-0042 USA; email ujsanchez@shockers.wichita.edu}
\thanks{A. Krishna Murthy is with the Department of Electrical Engineering and Computer Science, Wichita State University, Wichita, KS 67260-0042 USA; email axkrishnamurthy@shockers.wichita.edu}
\thanks{J. Steck is with the Department of Aerospace Engineering, Wichita State University, Wichita, KS 67260-0042 USA; email james.steck@wichita.edu}}%

\makeatletter
\newcommand{\newlineauthors}{%
  \end{@IEEEauthorhalign}\hfill\mbox{}\par
  \mbox{}\hfill\begin{@IEEEauthorhalign}
}
\makeatother
\begin{document}
\vspace{-50pt}
\maketitle

%\textit{Abstract}--
\vspace{-10pt}
\begin{abstract} A gate sequence of single-qubit transformations may be condensed into a single microwave pulse that maps a qubit from an initialized state directly into the desired state of the composite transformation. Here, machine learning is used to learn the parameterized values for a single driving pulse associated with a transformation of three sequential gate operations on a qubit. This implies that future quantum circuits may contain roughly a third of the number of single-qubit operations performed, greatly reducing the problems of noise and decoherence. There is a potential for even greater condensation and efficiency using the methods of quantum machine learning.
\end{abstract}
%\textit{Index terms}--

\section{Introduction}

Quantum computing is a field of growing interest with enormous potential; however, the current state of the art, called NISQ (Noisy Intermediate Scale Quantum), is limited due to problems with both fidelity and scaling. Methods for error correction have been devised, but they require a rapidly growing number of additional qubits, called ancilla qubits so that a smaller and smaller percentage of qubits can be used for the desired computation.

For several decades now, we have been exploring an alternative approach that has been demonstrated to be more robust to noise and decoherence: quantum machine learning (QML), rather than algorithm programming. In QML, a quantum system is used as a quantum computer that \textit{learns} how to perform a particular task when provided with an example dataset. In recent years, QML has emerged as a powerful tool for optimizing complex quantum systems. QML algorithms leverage the principles of quantum computing to process quantum data more efficiently than classical machine learning methods. Authors in \cite{Cerezo_2022} explore the intersection of quantum computing and machine learning, with a focus on the advantages of QML in accelerating data analysis, particularly, quantum data. Various QML models such as parameterized quantum circuits (PQCs) and Quantum Neural Networks (QNNs) are discussed while highlighting the potential for achieving quantum advantage in certain machine learning tasks, such as classification, optimization, and clustering. The study also throws light onto the challenges in QML, such as the trainability of quantum models, noise and decoherence in quantum hardware, the need for efficient quantum algorithms et cetra. \cite{Biamonte_2017} presents a theoretical framework and initial algorithms that demonstrate the significance for quantum speedup in machine learning. Further, in \cite{liang2022hybridgatepulsemodelvariational} both discrete gate-based and continuous pulse-based paradigms are combined producing a hybrid model for variational quantum algorithms (VQAs). The core objective in \cite{liang2022hybridgatepulsemodelvariational} is to improve the efficiency and accuracy of VQAs. 

In our current research, we employ QML to optimize microwave pulses used for controlling qubits in quantum computers. By enhancing the precision and effectiveness of quantum control mechanisms, our research focuses on improving the fidelity of quantum operations by using QML algorithms to train the parameters of the microwave pulses.  Prior work on pulse learning is in \cite{liang2022variationalquantumpulselearning}, where the authors propose variational Quantum Pulse (VQP) learning, to directly train quantum pulses for machine learning tasks. The research focused on evaluating the performance of VQP learning on binary classification tasks and the Modified National Institute of Standards and Technology (MNSIT) dataset. 

 In \cite{Melo_2023} the authors investigate the impact of pulse-efficient transpilation on near-term QML algorithm and cross-resonance-based hardware to improve the performance and fidelity of quantum circuits by reducing circuit schedule duration and mitigating the effects of device noise. Further, the authors introduce an approach to enhance the efficiency of QML via pulse-efficient techniques. While the study in \cite{Melo_2023} is more broadly focused on pulse efficiency across various QML tasks, our research specifically targets training microwave pulse for single qubit transformations. Moreover, our study emphasizes the condensation of gate sequences into a single microwave pulse. This potentially simplifies the implementations of specific qubit transformations.  
 
 The authors in \cite{meitei2021gatefreestatepreparationfast} propose an alternative algorithm called ctrl-VQE for state preparation in the variational quantum eigensolver (VQE) algorithm. While the results of this research show a reduction in state preparation times of roughly three orders of magnitude compared to gate-based strategies, there are challenges such as hardware limitations, optimization complexity, experimental implementation, scalability to larger systems et cetera, which need further investigation. \cite{schuld2021supervisedquantummachinelearning} presents the relationship between supervised quantum machine learning models and kernel methods, highlighting the mathematical similarities and implications for quantum models. 

 Additionally, in the research \cite{Xia_2018}, the authors present a hybrid quantum algorithm that employs a restricted Boltzmann machine in order to obtain molecular potential energy surfaces that are accurate. However, the main challenges lie in resource requirements. The simulations for the molecules used in their study, ($H_{2}$, $L_{i}H$ and $H_{2}O$) require 13 qubits each, leading to the simulation of simple molecules requiring a significant number of qubits. Further, noise management, scalability, implementation complexity et cetera, are some practical challenges that need to be addressed. 
 
 \cite{gianani2021experimentalquantumembeddingmachine} explores the idea of embedding classical data into a quantum state, which can then be manipulated in a larger Hilbert space. Optimization is carried out using gradient descent, employing automatic differential tools available in platforms such as PennyLane. These differential experimental platforms however come with unique constraints and challenges.  

In earlier work, we have proposed a general alternate method of designing algorithms by breaking them down into smaller “building blocks” of some universal set of gates.  We prepare the system at an initial time (the “input”), measure the system at a final time (the “output”), and use machine learning to train externally adjustable parameters to improve the closeness of the outputs to the desired values for the intended computation.
Apart from the reasons highlighted above regarding the importance of this study, further benefits are listed below, namely:

\begin{enumerate}
  \item It bypasses algorithm construction \cite{algdesign};
  \item It makes breaking down the computation unnecessary \cite{Thompson2020,geneticAlg,N-qubitSystemThompson};
  \item Scaleup is relatively easy \cite{multiqubitEntanglement}; and
  \item Multiple interconnectivity of the architecture results in robustness to both noise and decoherence \cite{Thompson2020, robustNoiseDecoherence, 2020paper}.
\end{enumerate}

In this paper, we extend our method to the level of the microwave pulses used to constitute the physical implementations of the quantum logic gates. Success suggests an increased efficiency for the overall computations, which allows the implementation of an entire transformation sequence -— represented as the product of its individual gates —- with far fewer pulses. Since the error rate of a quantum sequence grows with both the number of pulses and the time of operation, this would result in a lower error rate for a given calculation, and therefore, an increase in the length and depth of the possible computations.  This represents yet another advantage to the quantum machine learning approach.

\section{Methodology}
\begin{figure}[h]
    \centering
    \includegraphics[width=.55\linewidth]{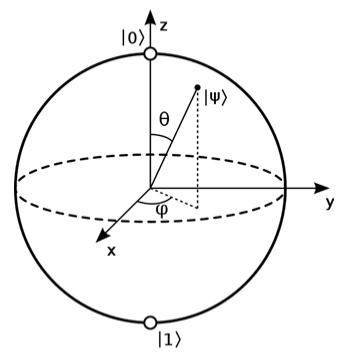}
    \caption{The Bloch sphere, whose poles correspond to the two charge basis states $\ket{0}$ and $\ket{1}$, and whose nonpolar regions represent superposition states. The surface of the sphere represents all possible pure states, given by Eq. \ref{purestate} above. For example, the state $\ket{0}$ is the point at the top of the depicted sphere with angles $\theta = \varphi = 0$; the equator represents any state in an equal superposition of $\ket{0}$ and $\ket{1}$, with $\theta = \pi/2$.}
    \label{BlochSphere}
\end{figure}

The pure state of a qubit, or quantum bit, can be generally written in the charge basis $\{\ket{0}, \ket{1}\}$ as
\begin{equation}\label{purestate}
    \ket{\psi} = \cos\left(\frac{\theta}{2}\right)\ket{0} + e^{i\varphi}\sin\left(\frac{\theta}{2}\right)\ket{1}
\end{equation}
where $0 < \theta < \pi$ and $0 < \varphi < 2\pi$ are angles on the Bloch sphere. See Figure \ref{BlochSphere}.

Single qubit operators move the qubit state from point to point on the Bloch sphere. Operators are called “gates” in analogy with classical logic gates, but in quantum computing, there are operators that cannot be expressed in terms of classical logic gates. For example, a Hadamard or $H$ operator acting on a qubit in state $\ket{0}$ or $\ket{1}$ results in a combined state of both $\ket{0}$ and $\ket{1}$ called a \textit{superposition}, which is impossible for classical bits to obtain. Superpositions are located on the nonpolar regions of the Bloch sphere. Operators can be represented on a given basis by a matrix, which transforms or maps one state onto another. For example, the Hadamard operator acting on the state $\ket{0}$ can be written as a matrix equation in the charge basis as shown in Equation \ref{hadamardgate}:
\begin{equation}\label{hadamardgate}
    H\ket{0} =
    \frac{1}{\sqrt{2}}
    \begin{pmatrix}
        1 & 1 \\
        1 & -1
    \end{pmatrix}
    \begin{pmatrix}
        1 \\
        0
    \end{pmatrix}
    = \frac{1}{\sqrt{2}}
    \begin{pmatrix}
        1 \\
        1
    \end{pmatrix}
\end{equation}
% The variables \vartheta, \chi, and \lambda are NOT the variables defined on the Bloch sphere figure (which is why vartheta and chi are used.)
All quantum gates are unitary operators, which conserve probability. In general, then, any single qubit gate is a mapping from one point on the Bloch sphere to another, which is conventionally written as a product of rotations around the x and z axes:
\begin{equation}\label{universalgate}
 U(\Delta\vartheta, \Delta\chi, \Delta\lambda) = R_z(\Delta\vartheta) R_x(\Delta\chi) R_z(\Delta\lambda) 
 \end{equation}
 This generalized rotation simplifies to the single matrix
 \begin{equation} 
    \begin{pmatrix}
        \cos\left(\frac{\Delta\vartheta}{2}\right) & -e^{i\Delta\lambda}\sin\left(\frac{\Delta\vartheta}{2}\right) \\
        e^{i\Delta\chi}\sin\left(\frac{\Delta\vartheta}{2}\right) & e^{i(\Delta\chi + \Delta\lambda)}\cos\left(\frac{\Delta\vartheta}{2}\right)
    \end{pmatrix}
\end{equation}
where $\Delta\lambda$, $\Delta\chi$, and $\Delta\vartheta$ are the angles of the rotations in Eq. \ref{universalgate}. For example, $U(\frac{\pi}{2}, 0, \pi)$ is the Hadamard gate ($H$); alternatively, $U(0, \pi, \Delta\lambda)$ is a rotation about the z-axis of the Bloch sphere by an angle $\Delta\lambda$, which simplifies to the $R_z(\Delta\lambda)$ gate.

\begin{figure}[h]
    \centering
    \includegraphics[width=\linewidth]{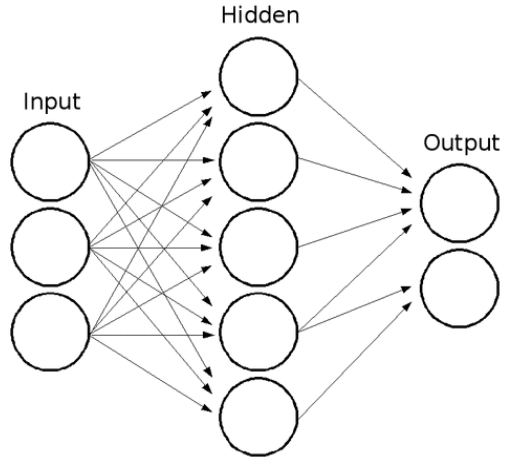}
    \caption{Diagram of a general neural network. The structure consists of an input layer where each circle represents a simple processing unit - a “node” or “neuron”. Each node in the input layer is connected to subsequent nodes in the “hidden layers” (here, there is only one hidden layer), finally connecting to the output layer on the right. In our current work, qubits represent the nodes, and the horizontal axis is time \cite{jadenFYRE}.}
    \label{fig:Neurons}
\end{figure}

% I think we could cut the paragraph below and Fig 2 and possibly Fig 3 if we need to save space Dr. Steck %
%\begin{comment} %added again for the journal - Anusha%
In previous work \cite{algdesign}, we have shown that a quantum system, adapted with machine learning and analogous to a neural network, can design its own algorithm, using the mathematical isomorphism between the time evolution of a quantum state and the information flow in a neural network. The isomorphism is as follows. A neural network consists of layers of nodes, with a typical feed-forward structure being one input layer, one or more hidden layers, and one output layer. See Fig. \ref{fig:Neurons}. We can think of this diagram of the information flow as analogous to time evolution by visualizing the horizontal dimension as ``time'', where, naturally, the system's state at any given time is causally connected to its state at previous times. The vertical dimension represents space: different parts of the system are connected to each other. 

%reworded to provide more clarification - Anusha 8/23/24 - Addressing the comments provided to the authors: (3)
In neural networks, the process by which a node (or neuron) produces an output based on its input and internal parameters is called "activating" the node. Quantum computing, however, does not have a direct one-on-one equivalent to the concept of "activation", but some analogous processes are available. We consider the key concepts and components of neural network activation and how they might map to quantum computing. Each node connects to every node of the next layer; each node computes a weighted sum of its inputs, in a neural network, which is then added to the bias term. This process is then followed up with passing the weighted sum through an activation function which helps determine the output of the node. In quantum computing, quantum algorithms make use of quantum gates for qubit manipulation. Quantum gates perform operations that alter the state of the qubits, similar to the activation function of a neural network. Similarly, in variational quantum algorithms (VQAs), rotational gates are parameterized akin to the activation function.

%Neural networks work to minimize a cost function (or performance index), which determines how wrong the output of the neural network is compared to what we want. %We adjust the weights and biases based on what would minimize the cost function. This corresponds to adjusting the (adjustable) parameters in the Hamiltonian for the quantum system. One pass-through of a training dataset of inputs and expected outputs is called an \textit{epoch}. 
% reworded Anusha 

Neural networks operate by minimizing a cost function (performance index), which quantifies the deviation of the network's output from the desired output. The weights and biases are adjusted to minimize the cost function, analogous to tuning the adjustable parameters in the Hamiltonian of a quantum system. A single pass through the training dataset, consisting of inputs and corresponding expected outputs, is referred to as an \textit{epoch}.

For our quantum system, the nodes are quantum mechanical in nature (i.e., each node is a qubit). The weights are the physical parameters of the quantum system, and the hidden layers are represented by various operations on quantum devices. 
%\end{comment} 

%According to the Schr\"{o}dinger equation, a pure quantum state evolves in time according to $\ket{\psi(t)} = e^{\frac{-i\hat{H}t}{\bar{h}}}\ket{\psi(0)}$
%Thus, the Hamiltonian $\hat{H}$ controls how the state evolves in time. The quantum system is prepared in a particular state (the “input”), allowed to evolve under a particular Hamiltonian $\hat{H}$, and finally measured (the “output”). The output of the measurement is compared to the desired output for that given input, and the Hamiltonian is then changed so as to decrease the error. The process is then repeated as shown in Fig. \ref{fig:Training}.
% reworded Anusha 

\begin{figure}[h]
    \centering
    \includegraphics[width=0.95\linewidth]{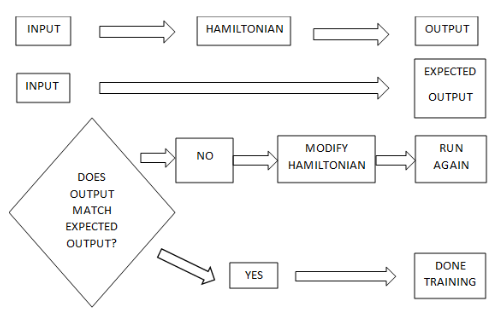}
    \caption{Process used to reduce final error when training. Training starts with a given input, a desired output, and an adjustable  Hamiltonian. The training system evolves in time according to a Hamiltonian, which can be altered to match the system's final state with the desired output.}
    \vspace{-7pt}
    \label{fig:Training}
\end{figure}

The training process is illustrated in Fig \ref{fig:Training}.
According to the Schr\"{o}dinger equation, a pure quantum state evolves in time according to $\ket{\psi(t)} = e^{\frac{-i\hat{H}t}{\bar{h}}}\ket{\psi(0)}$ .The Hamiltonian $\hat{H}$ governs the temporal evolution of the system's state. The quantum system is initially prepared in a specific state (block representing the `Input' in the figure below), undergoes evolution under the Hamiltonian $\hat{H}$, and is subsequently measured (block representing the `Output'). The measured output is compared to the desired outcome for the given input, and the Hamiltonian is adjusted to reduce the error, (block representing `Modify Hamiltonian'). This  process is iterated until the desired outcome is achieved, thus completing the training. 

This approach to programming quantum computers has been remarkably successful in both reproducing calculations done with other methods \cite{algdesign, Thompson2020} and in designing “algorithms” for which there is no known sequence of gates. This approach also has other advantages like scaleup and robustness to noise and decoherence \cite{multiqubitEntanglement, robustNoiseDecoherence}. More recently, the method has been rediscovered by others under the names of Quantum Approximate Optimization Algorithm (QAOA) \cite{turati2022, chandarana2022} and Variational Quantum Eigensolver (VQE) \cite{tao2022, nakayama2023vqe}. In some contexts, we are able to show that our method significantly simplifies and streamlines the desired computation by training the “whole thing” as opposed to breaking it down into building blocks \cite{algdesign, geneticAlg, Octavio2023}.  It is natural to ask if we can achieve even greater efficiency by applying machine learning at the sub-gate level.

\begin{figure}[h]
    \centering
    \includegraphics[scale=0.5]{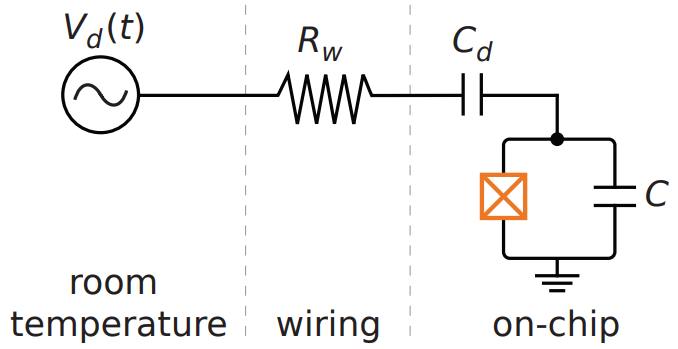}
    \caption{Circuit diagram for a transmon qubit attached to a microwave drive line \cite{krantz2019}.}
    \label{Transmon}
\end{figure}

Quantum states for superconductor qubits are controlled by microwave pulses characterized by a time-varying voltage $V_d(t)$ with specific parameters that result in a desired quantum transformation \cite{krantz2019}. See Fig. \ref{Transmon}.

\begin{figure}[h]
    \centering
    \includegraphics[width=1\linewidth]{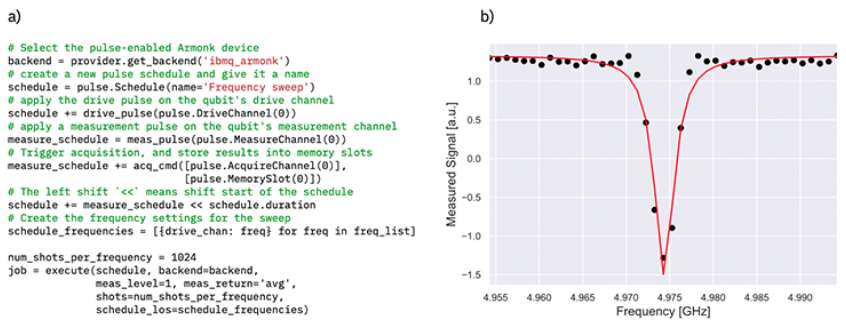}
    \caption{a) Commented code associated with running a job on IBM’s \textit{ibmq\_armonk} device. b) Graphically displays signal strength as a function of frequency to find the optimized frequency of a single qubit in the setup process \cite{asfaw2019get}.}
    \label{Resonance}
\end{figure}

In this paper, we use supervised machine learning to train these sub-gate pulses; therefore, OpenPulse permissions on the hardware are required to alter these microwave pulses. This is an uncommon feature for open-sourced quantum hardware devices. Most organizations with quantum hardware optimize and then fix the microwave pulse control frequency used in order to more closely match the resonant frequency of their system--shown in Fig. \ref{Resonance} below from IBM Research Blog’s post \cite{asfaw2019get}--which is why pulse-level alterations by users are not commonly allowed; however, IBM's \textit{ibmq\_armonk} device did allow users to implement pulse-level commands.

%In order to implement our trained pulses, we use the Qiskit Pulse \cite{ibmPulseQuantum, javadiabhari2024quantumcomputingqiskit} framework. Pulses are defined as  \textit{n} discrete time complex valued samples, $d_j$, $j \in \{0, 1, \dots, n -1\}$. As every quantum backend is different, we need to consider the time resolution available on the backend's processor, d\textit{t}, in our study, the sampling period, is 0.022ns. 
% Reworded Anusha 

Qiskit Pulse framework \cite{ibmPulseQuantum, javadiabhari2024quantumcomputingqiskit}, was used to implement our trained pulse. Pulses are defined as  \textit{n} discrete time complex valued samples, $d_j$, $j \in \{0, 1, \dots, n -1\}$. Various IBM quantum processors have different hardware specifications, including time resolution and sampling rate, details of which are not always explicitly published or uniform across all backends. These specifics can be found in the backend configuration data provided through IBM's Qiskit platform, based on which the time resolution d\textit{t} is considered. The sampling rate in our research is $\approx$ 45.45GHz. To achieve high measurement fidelity, the duration of the readout pulse must be longer than the time resolution d\textit{t} = 0.022ns. The small value of d\textit{t}  means that a readout pulse lasting for hundreds of nanoseconds would span thousands of d\textit{t}  units, leading to the readout pulse occupying over 20,000 d\textit{t} being typical for accurate quantum state measurements.

Each $d_j$ is then the value used for one timing cycle of the processor, also known as a timestep. At the  \textit{jth} timestep, the ideal value of the signal received by the drive channel, which relays the signal to the qubit \cite{Alexander_2020}, is:
\begin{equation}\label{pulsetimesample}
    D_j = \text{Re} \left[e^{2\pi fj \text{d}t + \phi} d_j\right]
\end{equation}
The modulation frequency \textit{f} and phase $\phi$
%WE USE A PHI HERE THAT IS DIFFERENT THAN THE PHI IN THE BLOCH SPHERE 
are properties of the drive channel and can be manipulated by the user through various instructions. For the rest of this work, it will be understood that "training a pulse" refers to using machine learning to train the complex values, $d_j$, which determine the microwave pulse.

\begin{figure}[h]
    \centering
    \includegraphics[width=1\linewidth]{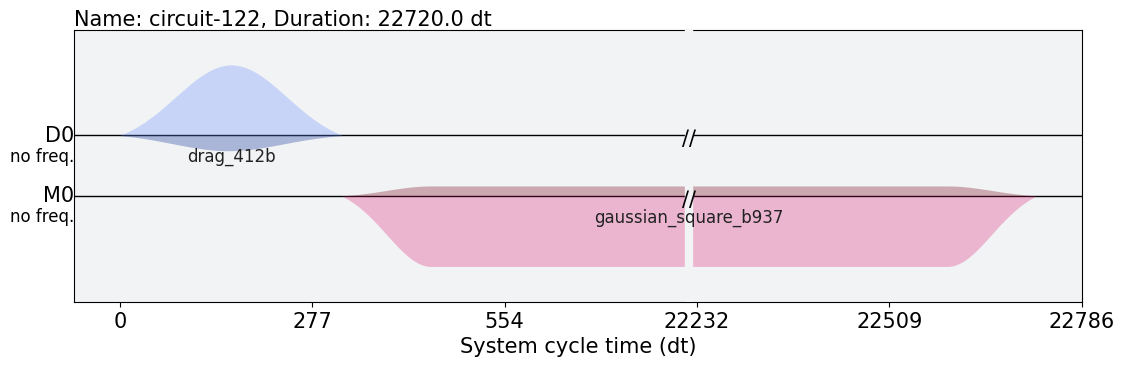}
    \caption{Pulses corresponding to an SX-gate and measurement. The symbol “D0” refers to the drive channel and “M0” refers to the measurement channel, both of which operate on the first qubit. This image was generated using the Python library Qiskit \cite{ibmPulseQuantum}.}
    \label{PulseRun}
\end{figure}

We train a single-qubit operation using the Derivative Removal by Adiabatic Gaussian (DRAG) pulse described in \cite{Gambetta_Motzoi_Merkel_Wilhelm_2011}. This involves training four parameters: duration, amplitude, variance, and correction amplitude. An example of such a pulse can be found in the drive channel D0 of the schedule shown in Fig. \ref{PulseRun}, while the pulse shown in the measurement channel M0 corresponds to the measurement of the qubit.

In our implementation, a ShiftPhase instruction will precede the DRAG pulse. The ShiftPhase instruction is responsible for modifying the phase of the drive channel, which results in an additional phase being added to all pulses played after the instruction. In other words, this instruction modifies $\phi$ in Equation \ref{pulsetimesample}. A ShiftPhase instruction is not necessary for training most of the gates shown in the results section below--such as the Pauli X and $\sqrt{X}$ gates (henceforth referred to as X and SX gates)--however, the Hadamard gate, $H$, did require this additional trained parameter.

Since the pulse parameters, $d_j$,  are complex, we train these as two separate real values: the signed modulus, \textit{r}, and the argument, $\alpha$, which are described in the following. We use the signed modulus instead of the traditional non-negative modulus to prevent discontinuities in the model. Furthermore, Qiskit Pulse prevents users from simulating a DRAG pulse whose magnitude is greater than one. Thus, instead of training the actual signed modulus, we train an input into Equation \ref{widesigmoid}, which maps our signed modulus into the interval~(-1,~1). 
\begin{equation}\label{widesigmoid}
    \sigma^*(x) = \frac{1-e^x}{1+e^x}
\end{equation}
\vspace{-0.25pt}
In this work, we define the effective signed modulus, $r_e$, and calculate the complex pulse parameter according to the formula $r_e e^{i\alpha} = \sigma^*(r)e^{i\alpha}$. If $r_e$ is positive, this form follows the standard exponential form for complex numbers. If $r_e$ is negative, this could instead be rewritten as $-r_e e^{i(\alpha + \pi)}$ in order to follow the standard convention. In all, this requires that we train the six parameters: duration, signed modulus, argument, variance, correction amplitude, and phase.

We implement our quantum neural network (QNN) using the Python libraries of Qiskit Dynamics (a supplementary library to Qiskit~\cite{ibmPulseQuantum}) and PyTorch~\cite{paszke2019pytorch}, with a single hidden layer. This hidden layer contains the six parameters described above and is the transformation defined as the qubit receiving a DRAG pulse with these parameters. We determine the final state of the qubit by using the pulse simulator from Qiskit Dynamics.

Training (input and target output) data was obtained by generating input states via equation \ref{purestate} with 10 random $\theta$ and $\varphi$ pairs.
%within their respective domains
The target outputs are the known resulting states after applying, to the input state, the known gate transformation we want to learn.  For example, the $X$ and $SX$ gates are well known and are given by Equations \ref{pauliX} and \ref{pauliSX}, respectively. 

\begin{center}
    \begin{tabular}{p{3cm}p{4.5cm}}
      \begin{equation}\label{pauliX}
    X = 
    \begin{pmatrix}
        0 & 1 \\
        1 & 0
    \end{pmatrix}
\end{equation}   & \begin{equation}\label{pauliSX}
    SX = \frac{1}{2}
    \begin{pmatrix}
        1+i & 1-i \\
        1-i & 1+i
    \end{pmatrix}
\end{equation} \\
    \end{tabular}
\end{center}
(Note that the $SX$ gate is, up to an overall phase, a rotation around the x-axis by $\frac{\pi}{2}$.) As such, we were able to operate on the state vectors of the initial states in the training dataset with these matrices in order to obtain our desired state after undergoing these transformations. 

A generic density matrix is defined by Equation \ref{densitymat}, 
\begin{equation}\label{densitymat}
    \sigma = \sum_i{{p_i}{\ket{\psi_i}\bra{\psi_i}}}
\end{equation}
where the sum is over all pure states $\ket{\psi}$ in the basis, and $p_i$ are the classical probabilities of each of the outer products so that $\sum_i{p_i} = 1$.  Here $\bra{\psi}$ is the Hermitian conjugate transpose  of $\ket{\psi}$. (Note that for a pure state, all but one $p_i$ are zero; the remaining $p_i$ is equal to one.) 

Our cost function for the machine learning algorithm is the infidelity (or error) between the target ($\sigma$) and actual output ($\rho$) density matrices given by Equation \ref{error}.
\begin{equation}\label{error}
    \text{Error} = 1 - F(\rho, \sigma) = 1 - \left( \text{tr}{\sqrt{\sqrt{\rho}\sigma\sqrt{\rho}}}\right)^2
\end{equation}

\noindent Here, \textit{tr} is the trace operator. It should be noted that fidelity has a range of zero to one, which is consistent with our definition of infidelity.  The \textit{fidelity} is a measure of the closeness of the final state to the desired state, so, we wish to minimize the \textit{infidelity} (our cost function).

\section{Results and Discussion}

We first trained some commonly used basis gates ($X$, $SX$, and $H$). The training was done in simulation using the Qiskit Dynamics \cite{qiskitcommunitySimulatingBackends, Puzzuoli:2023att} simulator with \textit{ibmq\_armonk} as a model. The training was rapid and complete for all single qubit gates attempted. Each job submitted for training consisted of 100 epochs. The gates trained are those listed in the first column of Table \ref{table1} at the top of the next page. The final error and the resulting pulse parameters for each gate are listed in the remaining columns of Table \ref{table1}. Errors (per Equation \ref{error}) were all on the order of $10^{-3}$ to $10^{-4}$.

\begin{table*}[tph]
\centering
\begin{tabularx}{1\textwidth}{| >{\centering\arraybackslash}X | >{\centering\arraybackslash}X | >{\centering\arraybackslash}X | >{\centering\arraybackslash}X | >{\centering\arraybackslash}X | >{\centering\arraybackslash}X | >{\centering\arraybackslash}X | >{\centering\arraybackslash}X | >{\centering\arraybackslash}X |}
 \hline
   Gate & Duration & Signed Modulus & Effective Signed Modulus & Argument & Variance & Correction Amplitude & Phase & Infidelity \\ \hline
 $X$ & 64.08 & 3.280 & 0.9275 & -0.5188 & 87.36 & -0.8577 & 0.06900 & 2.434E-3 \\ \hline 
 $SX$ & 77.29 & -1.167 & - 0.5254 & 0.3283 & 70.58 & -1.416 & -0.2190 & 1.097E-3 \\ \hline
 $H$ & 61.09 & 0.2154 & 0.1073 & -0.5127 & 86.51 & -1.294 & 2.010 & 2.360E-4\\ \hline
 $R_z$ & 55.23 & 1.070 & 0.4888 & -0.0253 & 70.86 & 0.7921 & 0.5244 & 1.521E-3\\ \hline
 $R_y$ & 67.26 & 2.606 & 0.8625 & -0.3393 & 72.96 & 0.7858 & -0.4742 & 3.160E-3\\ \hline
 $R_x$ & 63.37 & 1.204 & 0.5384 & -0.2262 & 72.22 & 0.6728 & -0.3898 & 4.744E-3\\ \hline
 U\_$R_x$ & 71.61 & 2.601 & 0.8618 & 0.7665 & 88.13 & 0.7125 & 0.3212 & 8.902E-3\\ \hline
 U\_$R_y$ & 51.87 & 2.977 & 0.9031 & -0.6832 & 70.66 & 0.7951 & -1.3309 & 2.757E-3\\ \hline
 U\_$R_z$ & 76.80 & 1.987 & 0.7589 & 0.3217 & 84.30 & 0.7418 & 0.1389 & 4.984E-3\\ \hline
 3-Rot & 79.77 & 2.491 & 0.8470 & -0.2740 & 77.36 & 0.6690 & 0.3204 & 3.702E-3\\ \hline
\end{tabularx}
 \caption{Trained single qubit gates showing trained parameters (duration, signed modulus, effective signed modulus, argument, sigma, beta, and phase) as well as infidelity (value of the cost function for the trained pulse.)}
 \vspace{-10pt}
 \label{table1}
\end{table*}

Next, we trained a ``combination'' gate, an example of the general single-qubit gate in Equation \ref{universalgate}. In previous work \cite{N-qubitSystemThompson}, we showed that an entanglement witness trained as a time-dependent Hamiltonian \cite{algdesign} could be rewritten as a sequence of gates, consisting of three rotations on a single qubit interspersed with a CNOT. We used this as an example to demonstrate consolidation.  Training of this gate is shown in the last entry in Table \ref{table1}: 3-Rot  $= R_y(-w_{5k+1}) R_z(w_{5k+1}) R_y(w_{5k+1})$, where $R_y$ and $R_z$ represent rotations about the $y$ and $z$ axes of the Bloch sphere respectively, and the rotation angle of $w_{5k+1}$ is determined through qubit biases and their tunneling amplitudes. Validity was again measured using Equation \ref{error}. Infidelity for this composite transformation gate was readily trained down to $3.702 \times 10^{-3}$  ($0.9963$ fidelity). This shows that our training methods can drive a qubit with one control pulse and produce an effect that would have traditionally required a plurality of pulses. This allows for the condensation of lengthy quantum circuits at little cost to error as well as the expedience of these circuits, as a single pulse is much faster and less error-prone than a multiplicity of pulses. 

\begin{figure}[h]
    \centering
    \includegraphics[width=0.97\linewidth]{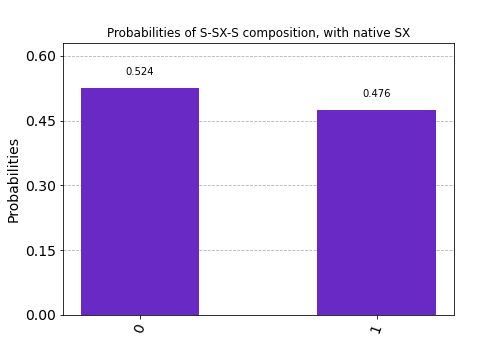}
     \vspace{-10pt}
    \caption{Bar graph of probabilities from an S-SX-S composition, with S being native gates on the computer and SX being our trained pulse \cite{jadenFYRE}.}
    \vspace{-7pt}
      \label{HComposition}
\end{figure}

\begin{figure}[h]
    \centering
    \includegraphics[width=0.98\linewidth]{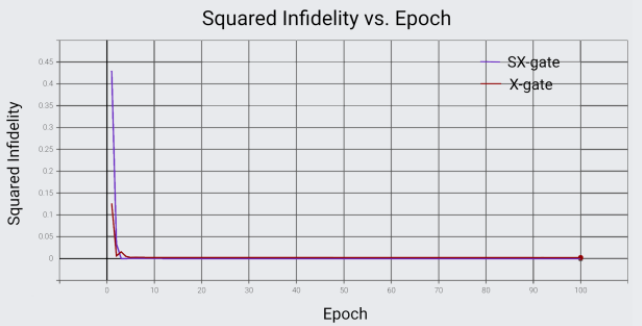}
    \vspace{-5pt}
    \caption{Error as a function of epoch for the SX and X gates 
    \cite{jadenFYRE}.}
    \vspace{-12pt}
    \label{InfVsEpoch}
\end{figure}

In addition to training on our simulator, for experimental verification, we tested gate identities on \textit{ibmq\_armonk}, a small-scale superconducting quantum processor that is accessible through the IBM Cloud.  One such verification process was by decomposing the Hadamard gate into an $S-SX-S$ composition and utilizing our trained $SX$-gate pulse parameters to juxtapose our methods with a known result. The resulting probabilities from our trained pulse is shown on the next page in Fig. \ref{HComposition}. These probabilities are nearly identical to the same gate composition, but with a native $SX$ gate on the computer, which gives us confidence in having trained a pulse with at most a difference in the global phase. A representative plot of error vs epoch for $X$ and $SX$ training is shown in Fig. \ref{InfVsEpoch}. 
\vspace{-10pt}

\hfill \break
The physical implementation of the transformations was challenging due to few accredited quantum computers being available that supported OpenPulse, an open-source framework developed by IBM that aids in designing and implementing custom quantum gate pulses. Further, OpenPulse is complex involving a steep learning curve for new users, error-prone, hardware-specific, and resource intensive, et cetera despite offering extensive quantum hardware control. In our future study, other alternatives such as IBM Quantum Experience, Qiskit and Quantum Scientific Computing Open User Testbed (QSCOUT) (by Sandia International Laboratories) will be made use of to compare and highlight the physical implementation of the transformations. The system associated with IBM’s Quantum Lab that we used, \textit{ibmq\_armonk}, retired \cite{RetiredSystems} during the neural network training. 

Our future work involves using IBM Quantum Eagle - a 127-qubit processor, and IBM Quantum Heron - a 133-qubit processor. As the developmental trends change in terms of having more reliable and powerful quantum computers that are open source, our experiments and findings shall also progress alongside it, providing us with ample opportunities to explore, research, and implement our approach further. In \cite{Ban_2023}, a QNN with multi-qubit interactions is proposed to improve the scalability of QNNs and minimize network depth without affecting the approximative power. The weight of the multi-qubit interactions term in the neural potential is updated using standard gradient descent. The QNN is trained for the task of searching prime numbers using a set of 2$n_b$ pairs for $n_b$ bits, where $n_b$ is the number of bits. To improve the research further, the scalability of QNNs with multi-qubit interactions can be investigated along with addressing potential challenges associated with the study. Additionally, exploring and developing specialized training algorithms tailored to QNNs with multi-qubit interactions can be done. 
\begin{comment}
Multi-qubit operations such as the CNOT and Toffoli could not be trained due to restrictions or proprietorship on information for how these were implemented; however, single qubit information was available.   --- Updated statement with new reference - Anusha  
\end{comment}
Our code is available at \cite{Github}.

\vspace{-2pt}
\section{Conclusion}
\begin{comment}
In previous work, we have shown that our method of quantum learning can be applied at the algorithm level \cite{algdesign} and the gate level \cite{Thompson2020}. 
\end{comment}
Quantum learning can be applied at the algorithm level \cite{algdesign} and the gate level \cite{Thompson2020} as shown in our previous work, here we extend our method to include the pulse level, leading to the reduction in the number of operations required to achieve certain states--which invariably would decrease the amount of error produced through the use of many different operations--as well as reduce the amount of time required to manipulate a qubit into those states, as each operation requires a microwave pulse of a certain duration to drive the qubit. 
\begin{comment}
 Currently, with lots of operations being applied, the amount of time our systems require to execute each operation significantly impairs the amount of time we are allowed to work with our qubit.    
\end{comment}
Through the use of our training methods, the amount of time required to execute each qubit operation can be dramatically reduced, allowing for a far greater number of multi-qubit operations to be applied than presently available. Future studies in this area could extend our work to determine the feasibility of condensing multi-qubit operations by making use of IBM Quantum Eagle and IBM Quantum Heron.
\vspace{-4pt}

\section{Acknowledgements}
\vspace{-0.5pt}
We thank the research group for their invaluable discussion in developing this project. We thank Wichita State University for partial support through the Undergraduate Research Experience (U. Sanchez) and First Year Research Experience(J. Nola). We are thankful to the various contributors of the Qiskit,  PyTorch library, and the IBM Quantum services for this work. 

\newpage

\end{document}